\documentclass{ragtime} 
\usepackage{physics}
\title[Magnetically Ejected Disks]%
      {Magnetically Ejected Disks: Equatorial Outflows Near Vertically Magnetized Black Hole}

\author[V. Karas, K. Sapountzis, \& A. Janiuk]   
       {Vladim\'{\i}r Karas\at[,]{1,a} 
         Kostas Sapountzis\at[,]{2} 
        \& Agnieszka Janiuk\at[]{2}\\\\   
       \ins{1}Astronomical Institute, Czech Academy of Sciences,  Bo\v{c}n\'{\i}~II 1401,\splitins[1] 
       CZ-14100~Prague, Czech Republic\\
       \ins{2}Center for Theoretical Physics, Polish Academy of Sciences, Al. Lotnikow 32/46,\splitins[1]
       P-02-668 Warsaw, Poland\\%
       \ins{a}\Email{E-mail:vladimir.karas@cuni.cz}} 
\begin{document}     
\newcommand{\apj}{ApJ}
\newcommand{\apjs}{ApJS}
\newcommand{\apjl}{ApJL}
\newcommand{\pasp}{PASP}
\newcommand{\mnras}{MNRAS}
\newcommand{\aj}{AJ}
\newcommand{\nat}{Nature}
\newcommand{\nar}{NewAR}
\newcommand{\na}{NewA}
\newcommand{\aap}{A\&A}
\newcommand{\aaps}{A\&AS}
\newcommand{\icarus}{Icar}
\newcommand{\araa}{ARA\&A}
\newcommand{\aapr}{A\&ARv}
\newcommand{\aplett}{Astrophysical Letters}
\newcommand{\prd}{Phys. Rev. D}
\newcommand{\actaa}{Acta Astronomica}

\setcounter{page}{107}

\begin{abstract}
Black holes attract gaseous material from the surrounding environment. Cosmic plasma is largely ionized and magnetized because of electric currents flowing in the highly conductive environment near black holes; the process of accretion then carries the magnetic flux onto the event horizon, $r\simeq R_+$. On the other hand, magnetic pressure acts against accretion. It can not only arrest the inflow but it can even push the plasma away from the black hole if the magnetic repulsion prevails. The black hole does not hold the magnetic field by itself. 

In this contribution we show an example of an equatorial outflow driven by a large scale magnetic field. We initiate our computations with a spherically symmetric distribution of gas, which flows onto the domain from a large distance, $r\gg R_+$. After the flow settles in a steady (Bondi) solution, we impose an axially symmetric configuration of a uniform (Wald) magnetic field aligned with the rotation axis of the black hole. Then we evolve the initial configuration numerically by employing the MHD code that approaches the force-free limit of a perfectly conducting fluid. 

We observe how the magnetic lines of force start accreting with the plasma while an equatorial intermittent outflow develops and goes on ejecting some material  away from the black hole. 
\end{abstract}


\section{Introduction}\label{introduction}
Accretion is ubiquitous process in the Universe. By gradual accretion, various components of highly diluted environment are gathered and brought onto cosmic bodies -- planets, stars, even galaxies \citep[see, e.g.,][]{2017ASSL..430.....F,2018ASSL..454.....S}. Accretion is driven primarily by action of gravitational and electromagnetic forces. Gravity of the central body acts at long range and it attracts gas and dust, whereas the electric and magnetic components can act in a complex, mutually interrelated manner; they can cause either attraction or repulsion of the plasma, depending on the conditions.

Astrophysical plasmas are characterized by their high conductivity which ensures that the force-free condition is typically an excellent approximation to describe the cosmic environment \citep{2010mfca.book.....B}. Moreover, in magnetospheres of compact objects like neutron stars and black holes there are regions where the energy density of the electromagnetic field greatly exceeds the inertial (kinetic, rest-mass, and thermal) energy of matter.  Plasma motions follow the evolving magnetic field lines. Force-free electrodynamics describes magnetically dominated relativistic plasma as long as the inertial forces can be neglected; while this assumption is correct, in very diverse circumstances it becomes gradually violated in case of very low ionisation and low temperature, where the dissipation effects play a role, and for ultra-relativistic acceleration near the light cylinder, where the particle mass is important.

In the limit of vanishing magnetic field and sufficiently high density (short mean free path) the hydrodynamical approach is adequate. The best-known analytical framework then describes the stationary, spherically symmetric inflow, a.k.a. Bondi solution \citep{1952MNRAS.112..195B}, where the actual form of the flow is determined by the boundary conditions at infinity and at the black hole horizon. This has been generalized in several ways; in particular, \citet{2008ApJ...686..172S} include the effect of additional source of energy from stars of the Nuclear Star Cluster, which is relevant for many nuclei containing Nuclear Star Clusters, including the Galactic center source Sgr A* \citep{2014A&A...566A..47S}. On the other hand, in the limiting case of zero density (electro-vacuum) the solution is described by the source-free coupled Einstein-Maxwell equations. These are tractable only under very constraining assumptions and symmetries, however, the problem can be simplified for electro-magnetic fields that are weak (albeit non-vanishing) in comparison with the gravitational field (the assumption valid in the vicinity of astrophysical black holes). Electro-magnetic test field solutions on the fixed background of Kerr metric then provide an adequate description of magnetic fields in interaction with the black hole gravity \citep{1974PhRvD..10.1680W,1975PhRvD..12.3037K}. 

In this contribution we are interested in a gradually evolving structure of magnetic field, as the system goes over from the initial, homogeneous solution to the interaction with the force-free magnetosphere near an accreting Kerr black hole \citep{2002apa..book.....F,2008bhad.book.....K}. We deliberately impose axial symmetry along the black hole rotation axis for the magnetic field and the inflowing medium (we employ a two-dimensional scheme). Although this constraint will have to be relaxed to describe astrophysically realistic systems, we want to reveal the transition from the initial state of magnetic lines running around the black hole, i.e., the magnetic flux being expelled out of the horizon \citep[partially in the case of non-rotating or moderately rotating black hole, dimension-less spin parameter $|a|<1$, and completely in the case of extreme rotation, $a=1$; see][]{1976GReGr...7..959B}. As the force-free plasma starts inflowing with spherical symmetry at the initial configuration, the Meissner expulsion is immediately diminished and, at later stages, the magnetic lines start to produce reconnection regions in the equatorial plane. Subsequently, localized blobs emerge and they are eventually ejected away from the black hole due to the magnetic pressure.

We build our study following a series of works that have been previously published by various authors. In particular, \citet{2007MNRAS.377L..49K} explored the magnetized, rotating black holes embedded in the plasma. By employing the general-relativistic magneto-hydrodynamics 2D HARM code \citep{2003ApJ...589..444G,2006MNRAS.368.1561M,2019ApJ...873...12S} they found that the Meissner expulsion indeed disappears due to the presence of accreting medium even in the case of (almost) maximally rotating black hole. \citet{2014PhRvD..89j4057P} gave arguments to understand the essence of the Meissner--type magnetic field expulsion near black holes. Various authors \citep{2016ApJ...816...77P,2016PhLB..760..112G,2020arXiv200715662C} studied the analytical properties of force-free black-hole magnetospheres especially in the context of jets emerging from the vicinity of the ergosphere. \citet{2014ApJ...788..186N} and \citet{2018PhRvD..98b3008E} noticed the formation of current sheets near a black hole immersed in a magnetized plasma. Some results suggest that the role of ergosphere is essential in producing the plasma structures and ejecting matter in the force-free medium \citep{1977MNRAS.179..433B}.

\section{The model set-up}
\subsection{Strong gravitational and weak magnetic fields}
Gravitational field is described by Kerr metric, which can be written in the well-known Boyer-Lindquist coordinate system $(t,r,\theta,\phi)$ \citep{1973grav.book.....M,1983mtbh.book.....C,1984ucp..book.....W}. This spacetime is asymptotically flat and it obeys the axial symmetry about the rotation axis and stationarity with respect to time; the singularity is hidden below the event horizon. The mass $M$ of the black hole is concentrated in the origin of the coordinate system. The spin parameter $a$ of the Kerr metric describes its rotation; the condition about the presence of the outer event horizon at a certain radius, $r=R_+$ (where the horizon encompasses the singularity) leads to the maximum value of the dimensionless spin rate: $|a\leq1|$. The solution can be then written in the form of the metric element \citep{1973grav.book.....M,1983mtbh.book.....C}
\begin{equation}
{\rm d}s^{2} = 
  -\frac{\Delta\Sigma}{A}\,{\rm d}t^{2}
   +\frac{\Sigma}{\Delta}\,{\rm d}r^{2}
   +\Sigma\,{\rm d}\theta^{2}
+\frac{A\sin^2\theta}{\Sigma}\;
   \left({\rm d}\phi-\omega\,{\rm d}t\right)^{2},
\label{metric}
\end{equation}
where $\Delta(r)=r^{2}-2r+a^{2}$, $R_+=1+\sqrt{1-a^2}$, $\Sigma(r,\theta)=r^{2}+a^{2}\cos^2\theta$, $A(r,\theta)=(r^{2}+a^{2})^{2}-{\Delta}a^{2}\sin^2\theta$, $\omega(r,\theta)=2ar/A(r,\theta)$. Dimension-less geometrical units are assumed with the speed of light $c$ and gravitational constant $G$ set to unity. In physical units the gravitational radius is thus equal to $R_{\rm{}g}=c^{-2}GM\approx4.8\times10^{-7}M_7\,$pc; the corresponding light-crossing time-scale $t_{\rm{}g}=c^{-3}GM\approx49\,M_7\,$sec, where $M_7\equiv M/(10^7M_\odot)$. 

Kerr metric is a solution of Einstein's equation for the gravitational field of a rotating black hole in vacuum. Even in the case of strongly magnetized gaseous environment around the black hole the contribution of an astrophysically realistic magnetic energy to the space/time curvature is negligible. We can thus neglect its effect on the metric terms and assume a weak-field limit on the background of Kerr black hole; the space-time metric is not evolved in our scheme. In order to initiate the numerical code we can employ an initially uniform magnetic field \citep{1974PhRvD..10.1680W,2007IAUS..238..139B}, which is fully described by two non-vanishing components of the four-potential,
\begin{eqnarray}
A_{t} &=& Ba\Big[r\Sigma^{-1}\left(1+\cos^2\theta\right) -1\Big], \label{mf1}\\
A_{\phi} &=& B\Big[{\textstyle\frac{1}{2}}\big(r^2+a^2\big)
 -a^2r\Sigma^{-1}\big(1+\cos^2\theta\big)\Big] \sin^2\theta \label{mf2},
\end{eqnarray}
in dimension-less Boyer-Lindquist coordinates and $B$ is the magnetic intensity of the uniform field far from the event horizon. The magnetic field (and the associated electric component) are generated by currents flowing in the accreted medium far from the black hole, as the latter does not support its own magnetic field. The set of two non-vanishing four-potential vector components defines the structure of the electromagnetic tensor, $F_{\mu\nu}\equiv A_{[\mu,\nu]}$; by projecting onto a local observer frame one then obtains the electric and magnetic vectors $\vb*{E}$ and $\vb*{B}$.\footnote{Let us note that typical cosmic plasmas are ionized and perfectly conducting, and so the approximation of force-free electromagnetic action is justified at high accuracy \citep[e.g.][]{2012ASSL..391.....S}: $\vb*{E}+\vb*{v\times B}=0$, where $\vb*{E}$ and $\vb*{B}$ are electric and magnetic intensities, $\vb*{v}$ is velocity of the plasma. Magnetic field is thus frozen in the plasma.}  However idealized the initial configuration may be, the numerical solution rapidly evolves in a complex entangled structure, with field lines turbulent within the accreting medium and more organized in the empty funnels that develop outside the fluid structure.

\subsection{Two limiting cases for the initial distribution of plasma}
The plasma forms an accretion disk or a torus residing in the equatorial plane, so that the axial symmetry is maintained. In a non-magnetized (purely hydrodynamical) limiting case one can find the classical solution for the density distribution $\rho\equiv\rho(r,z)$ and pressure $P\equiv P(r,z)$, and the geometrical shape $H\equiv H(r)$ of a non-gravitating barytropic torus $P=K\rho^k$. Introducing enthalpy of the medium, $W(P)\equiv\int dP/\rho$ and setting $P_{\rm in}=P_{\rm out}=0$ at the inner and the outer edges of the density distribution \citep{1978A&A....63..221A,1978A&A....63..209K}
\begin{equation}
 W_{\rm out}-W_{\rm in}=\int_{R_{\rm in}}^{R_{\rm out}}
  \,\frac{l(R)^2-l_{\rm kep}(R)^2}{R^3}\,dR=0,
\end{equation}
where $l_{\rm kep}(R)$ is the radial profile of the Keplerian angular momentum density in the equatorial plane. Several properties of this solution are worth mentioning  \citep{1971AcA....21...81A}: (i)~The level surfaces of functions $P,$ $\rho,$ and $W$ coincide; (ii)~If the torus boundary $P=0$ forms a closed surface, the torus center is defined by where $dP/dR=0,$  the pressure is maximum; (iii)~The shape of the torus can be found by integrating the vertical component of the Euler equation.

Above a certain critical value, $W>W_c,$ the torus forms a stable configuration (see the shaded region), while for $W<W_c$ matter overflows onto the central object even if we neglect viscosity. This behaviour resembles the Roche lobe overflow in binary systems, however, here it is a consequence of the non-monotonic radial dependence of the Keplerian angular momentum in the relativistic regime near the black hole \citep{1980ApJ...242..772A,2013A&A...559A.116P}. Whereas the material inflowing from atmosphere of  a primary component of the binary system concentrates near the equatorial plane and naturally forms the torus, in the case of a single, isolated central black hole matching the inner (toroidal) structure to the outer reservoir of matter depends on many circumstances at the outer boundary region. This is also the case of super-masive black holes residing in nuclei of galaxies, which are fed by interstellar medium from a surrounding (spheroidal) nuclear star cluster and the galaxy bulge; flattening of the structure is a parameter that can vary from disk-type equatorial inflow up to perfectly spherical (Bondi-type) inflow/outflow solution \citep{2008ApJ...686..172S,2017MNRAS.464.2090R}.

Unlike the above-discussed toroidal configuration, let us now assume that the angular momentum of accreted fluid is negligible and its velocity has a non-vanishing component only in the radial direction. At large radius, in our case suitable as a boundary condition, we are allowed to consider the problem within the framework of spherically symmetric Newtonian inflow with $v_r=v<0$ (positive $v$ would correspond to a symmetrical problem of an outflow or a wind). The equation of continuity gives $4\pi r^2\rho(r) v(r)=-\dot{M}$, where the constant on the right-hand side has a meaning of mass accretion rate. In the Euler equation, density of the external force $\vb*{f}$ has
only the radial component, $GM\rho/r^2=-f_r(r)$, and so we can write
\begin{equation}
 v\,\frac{dv}{dr}+\frac{1}{\rho}\frac{dP}{dr}+\frac{GM}{r^2}=0.
 \label{eul1}
\end{equation}
Introducing the sound speed by $dP=c_{\rm s}^2(r)\,d\rho$ the Euler equation can be manipulated into the well known form
\begin{equation}
 \frac{1}{2}\left(1-\frac{c_{\rm s}^2}{v^2}\right)\frac{d\,v^2}{dr}=
 -\frac{GM}{r^2}\left[1-\frac{2c_{\rm s}^2r}{GM}\right].
 \label{eul2}
\end{equation}
The solutions can be classified according to their behaviour at the sonic point, where the medium flows at the speed of sound, $r_{\rm s}=GM/2c_{\rm s}^2(r_{\rm s})$. For the spherical adiabatic accretion one can find six qualitatively different solutions to the above equations in the $(v,r)$-plane. Inflows (accretion flows) and outflows
(ejection or stellar wind) are both possible; which mode is realized in a particular situation depends on boundary conditions (in our case, only accretion is possible at the inner boundary).

\begin{figure}[tbh!]
 \centering
 \includegraphics[width=0.65\textwidth]{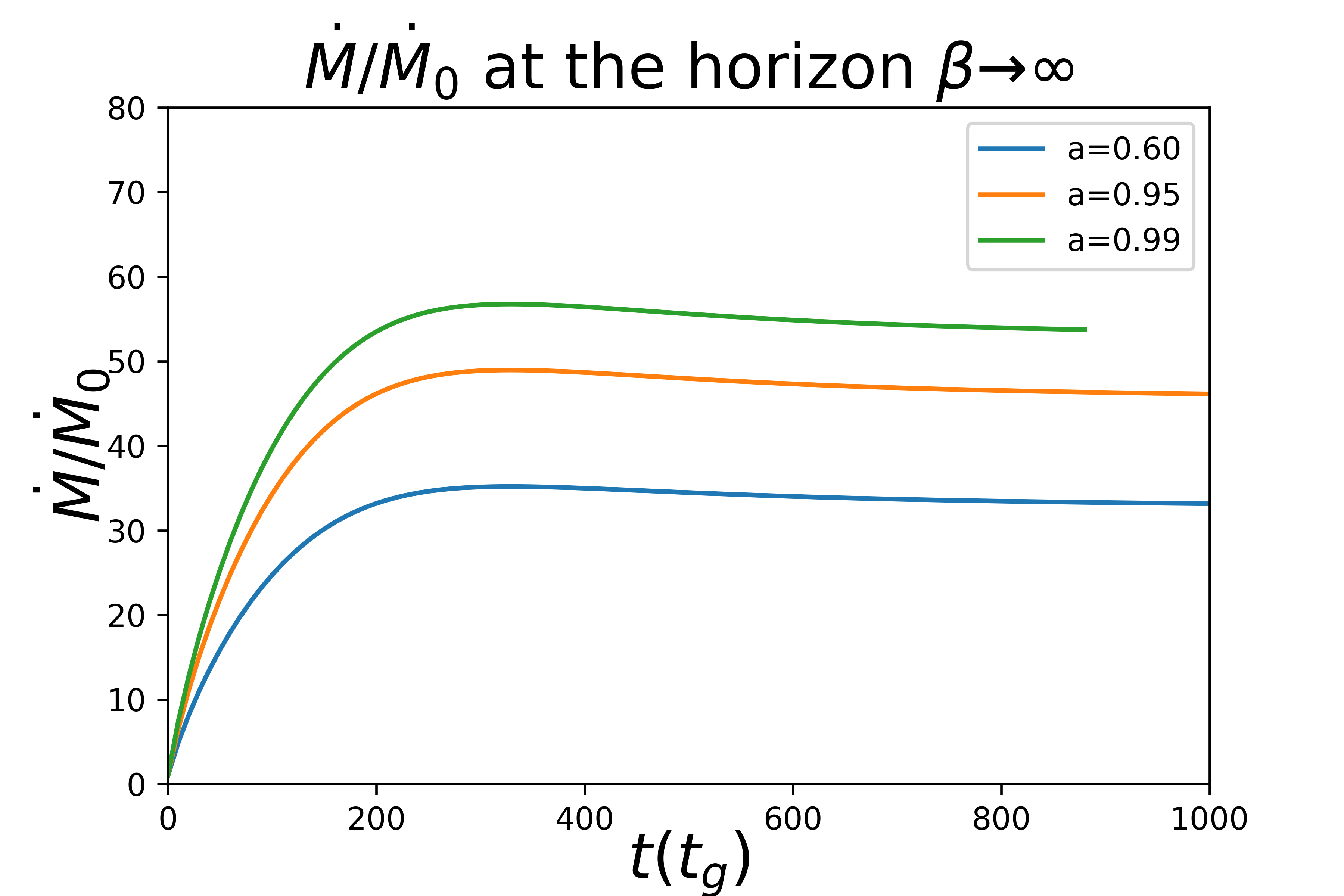}
 \caption{The accretion rate on the horizon of the black hole initially grows and then saturates as the steady state has been reached. The curves are normalized to an arbitrary value $\dot{M_0}$ and parameterized by three values of the dimension-less spin $a$. Time is in geometrical units and scaled with the black hole mass $M$.}
 \label{fig1}
\end{figure}

Let us note that the assumptions about constancy of $l(r)$ and the vanishing magnetization, $\beta=\infty$, are astrophysically unrealistic but they are useful to simplify calculations and allow an analytical insight. The magnetization $\beta$ parameter is taken here as ratio of total hydrodynamic pressure to the magnetic pressure within the magnetized fluid, i.e., $\beta\equiv U_{\rm tot}/U_{\rm mag}$. Realistic models must relax the extreme assumptions about the strict geometrical symmetry and stationarity to allow the system to evolve in time, which is a crucial aspect of the mutual interaction between different components. 

As mentioned above, to overcome some of the limitations we adopt the numerical scheme by the HARM code \citep{2003ApJ...589..444G,2006MNRAS.368.1561M}.\footnote{In this paper we explore an axially symmetric configuration, which is obeyed by all components of the system: the gravitational field of the rotating black hole, the interacting electromagnetic field of external origin, and the fluid surrounding the black hole. We thus employ the two-dimensional version of the code. Imposing the axial symmetry allows us to examine the role of magnetic expulsion from the horizon of extreme Kerr black hole and to observe how this effect is reduced by the accreted plasma. At the same time we are not confused by non-axisymmetric effects, which are known to reduce the magnetic expulsion, too. It will be interesting to generalize our discussion to a non-axisymmetric configuration of an oblique (inclined with respect to the rotation axis) magnetic field.} This allows us to explore the parameter space of the system. A fraction of the material injected spherically at the outer boundary of the computational domain remains bound and it forms an accretion torus or an accretion disk associated with the central black hole. This occurs at a relatively small radius where relativistic effects play a crucial role and decide whether the gas falls onto the black hole or becomes redistributed and ejected (we do not include radiative cooling in the present work). The interplay between the frozen-in magnetic field and the infalling plasma that thermalizes the mechanical energy and generates additional overpressure, eventually determines the ratio between accretion and ejection.

\begin{figure}[tbh!]
 \centering
 \includegraphics[width=0.48\textwidth]{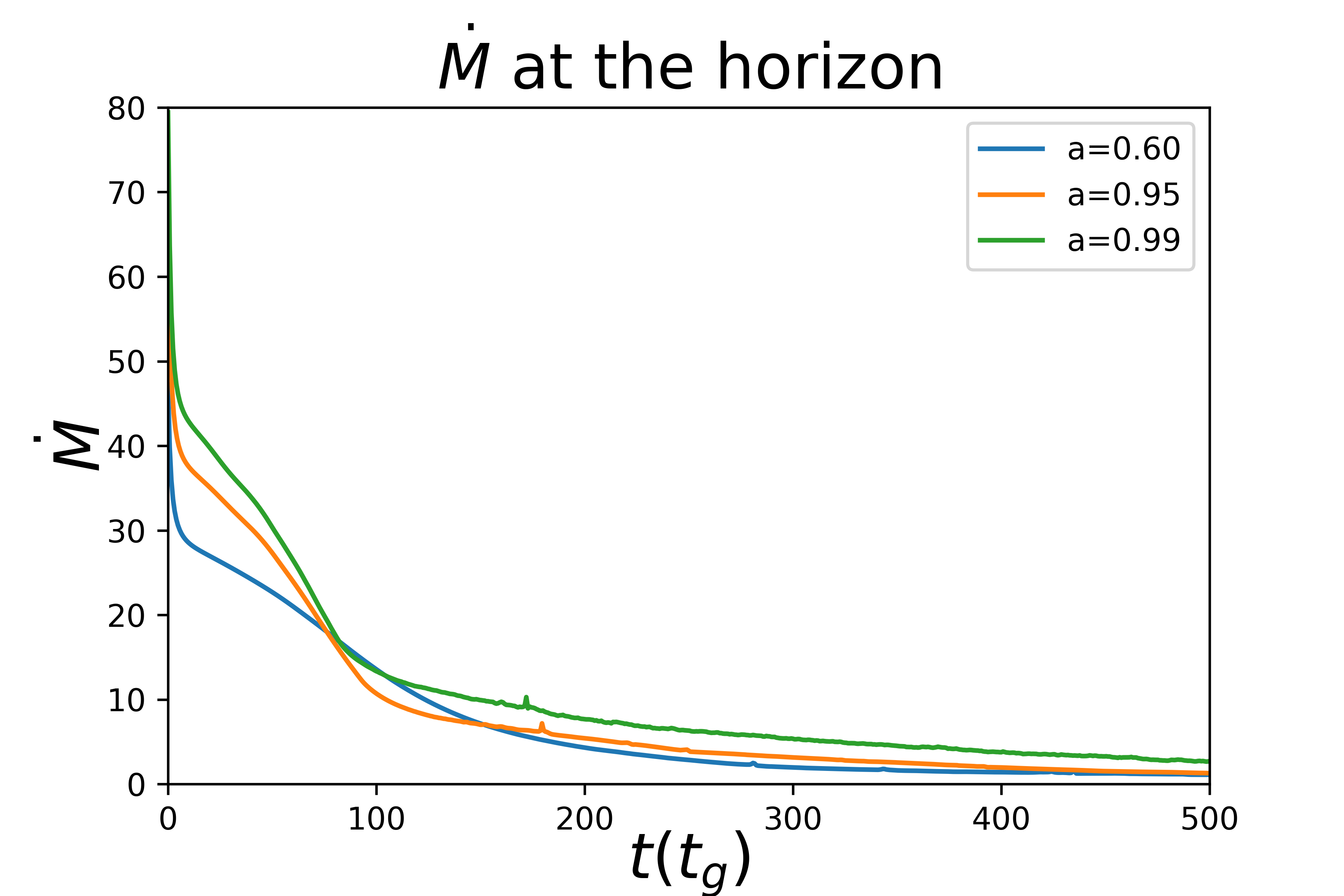}
 \includegraphics[width=0.48\textwidth]{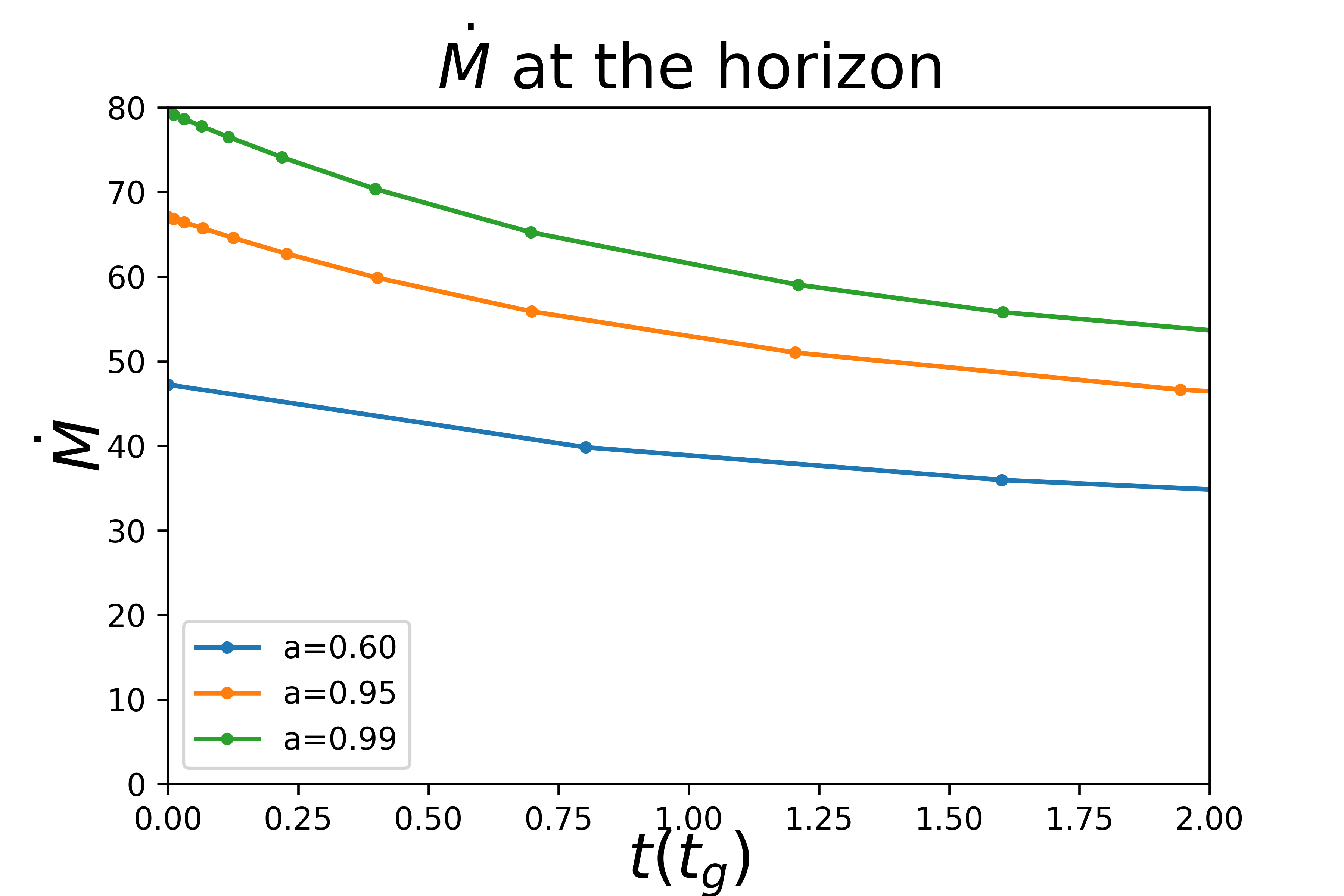}
 \caption{Gradually evolving accretion rate onto the magnetize black hole with the initial set-up of the Bondi inflow and the imposed Wald magnetic field for three values of spin $a$. The magnetization parameter $\beta\simeq0.1$ (magnetically dominated medium) at $t=0$. Long-term progress is shown in the left panel; a detail of the initial phase in the right panel.}
 \label{fig2}
\end{figure}

\section{Results}
\subsection{Mass accretion rate}
We set the outer boundary of the computational domain at the radius $10^3R_{\rm g}$, where the inflow is purely radial at initial time. The inner boundary is set at $\simeq0.65R_{\rm g}$, i.e.\ a fraction of gravitational radius and also inside the horizon radius for the corresponding value of spin $a$. The grid domain has been resolved at $600\times512$ points in $(r,\theta)$ coordinates and the polytropic index set to $k=4/3$. We use the magnetic intensity $B$ of the Wald field to determine the plasma parameter $\beta$ (in our notation, $\beta$ closer to zero corresponds to a more magnetized plasma). 

In order to initialize the computation we employ the hydrodynamic (non-mag\-net\-ized), purely spherical inflow. We set $\beta\rightarrow\infty$ and allow the inflow to build a steady-state Bondi accretion at a certain level of $\dot{M}$ (see Fig.~\ref{fig1}). Once the inflow stabilizes to a quasi-steady state inflow, we impose the large-scale Wald magnetic field along the rotation axis, which is then evolved further. Because of the perfect conductivity and the force-free approximation (apart from the effective small-scale numerical dissipation), the magnetic field-lines remain attached to plasma. However, the evolution of the system can be strongly altered if the magnetic field is strong enough, so that its repulsive tendency halts accretion. This effect is governed by the magnetization $\beta$-parameter, which is not uniform across the computational domain and changes in time. While the accretion is not much influenced in the limit of negligible magnetization ($\beta\gg1$), where the gravitational attraction of the black hole prevails, in the case of equipartition between the magnetic and hydrodynamic pressure ($\beta\approx1$) near the horizon the inflow is partially diverted into an outflow and the accretion rate is diminished (see Fig.~\ref{fig2}). Let us note that the mass and spin of the black hole are not updated during the simulation because the amount of accreted material is tiny compared to the black hole mass.

\begin{figure}[tbh!]
 \centering
 \includegraphics[width=0.65\textwidth]{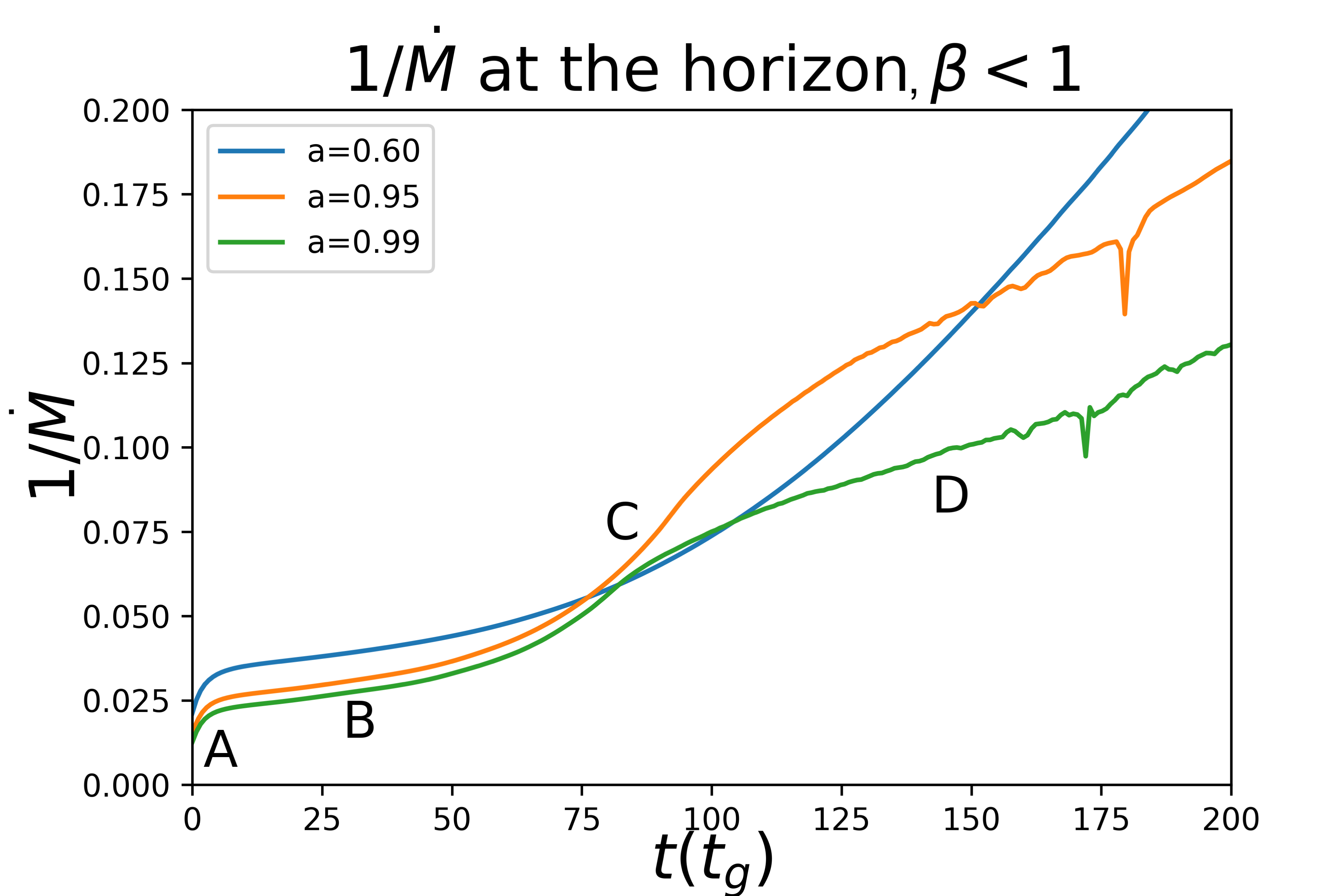}
 \caption{Graph of inverse accretion rate $1/\dot{M}$ exhibits four phases of the magnetized flow evolution. The initial configuration starts with the Bondi spherical inflow from the outer boundary of the integration domain. The central black hole rotates with the Kerr spin parameter $a$ and it is magnetized by the Wald uniform magnetic field $B$. Spherical symmetry of the inflow is quickly lost by its interaction with the magnetic field but the axial symmetry is imposed in our 2D computations. In case of a rapidly rotating black hole, the magnetic flux vanishes initially (Meissner effect) but it starts growing with accretion of the plasma. Part of the inflowing material is diverted to an outflow along the equatorial plane and accelerated by reconnection events (they are caused by numerical resistivity); the resulting accretion rate thus gradually drops and it exhibits some random glitches at later stages of its temporal evolution (see the text for further details).}
 \label{fig3}
\end{figure}

\begin{figure}[tbh!]
 \centering
 \includegraphics[width=0.65\textwidth]{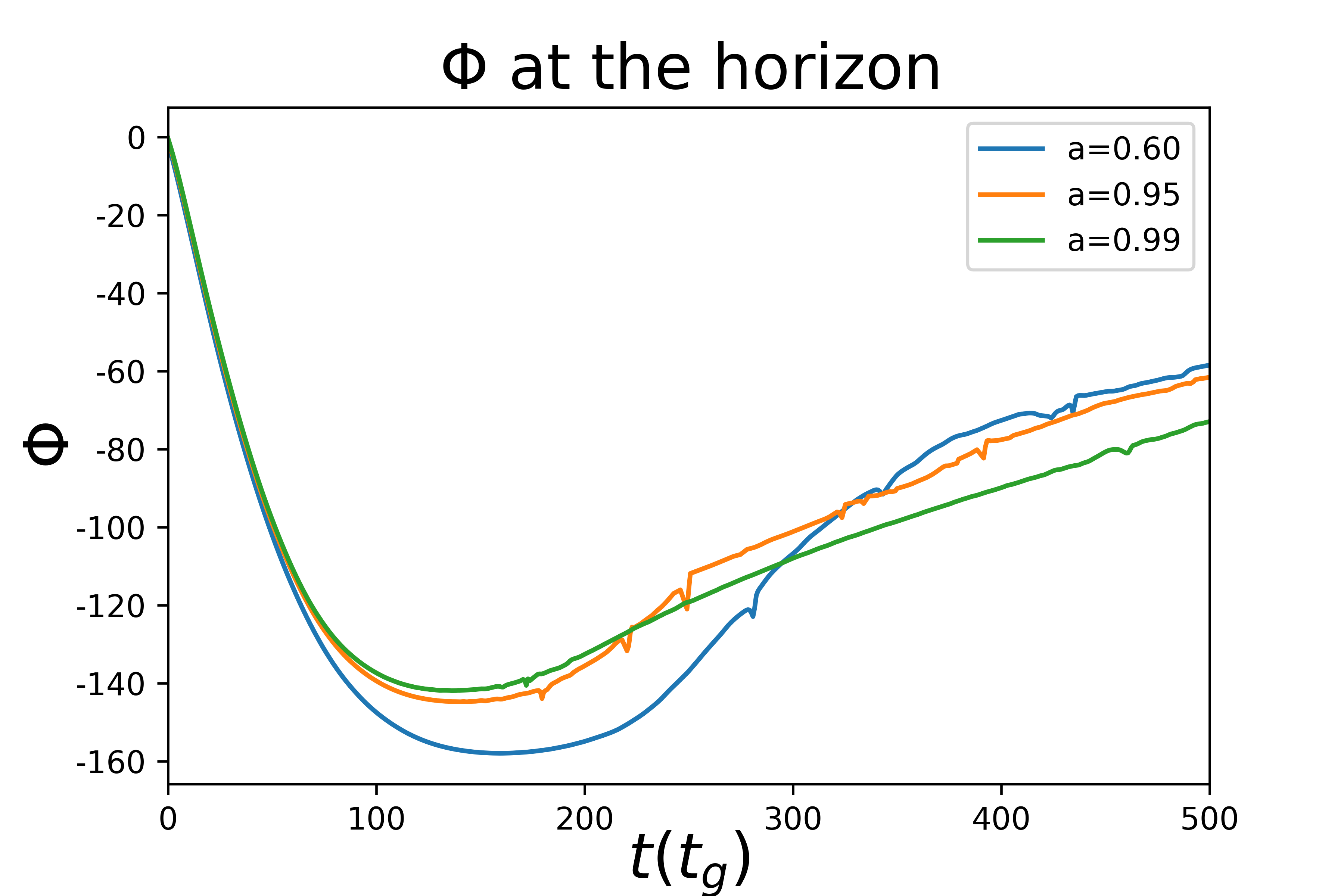}
 \caption{The magnetic flux $\Phi(t)$ (in arbitrary code units) across a hemisphere on the black hole horizon. The initially frozen-in magnetic flux grows (in absolute value) due to accretion of plasma. At later stages the flux starts decreasing as the magnetic intensity decreases and the field eventually escapes to radial infinity.}
 \label{fig4}
\end{figure}

\begin{figure}[tbh!]
 \centering
 \includegraphics[width=0.48\textwidth]{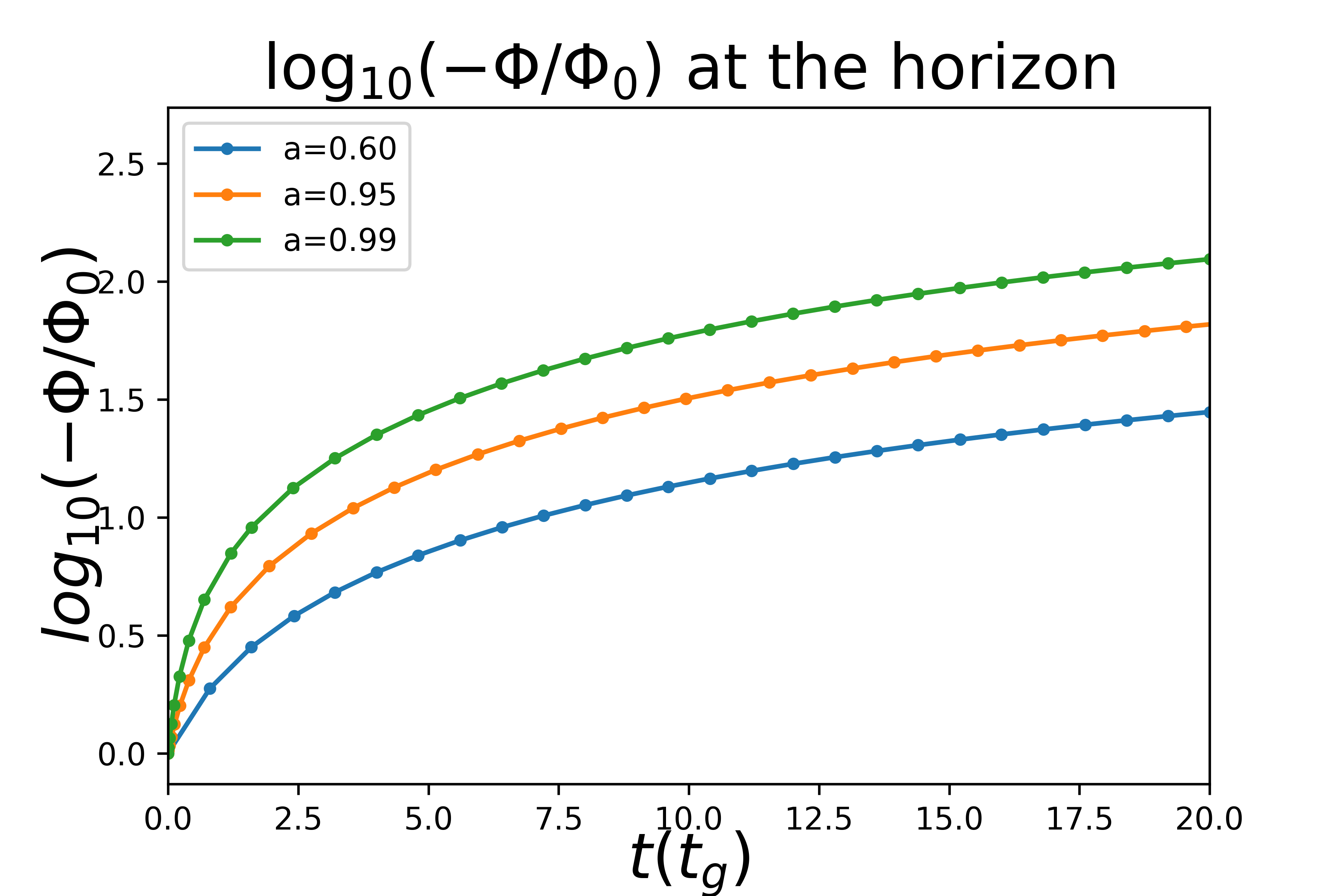}
 \includegraphics[width=0.48\textwidth]{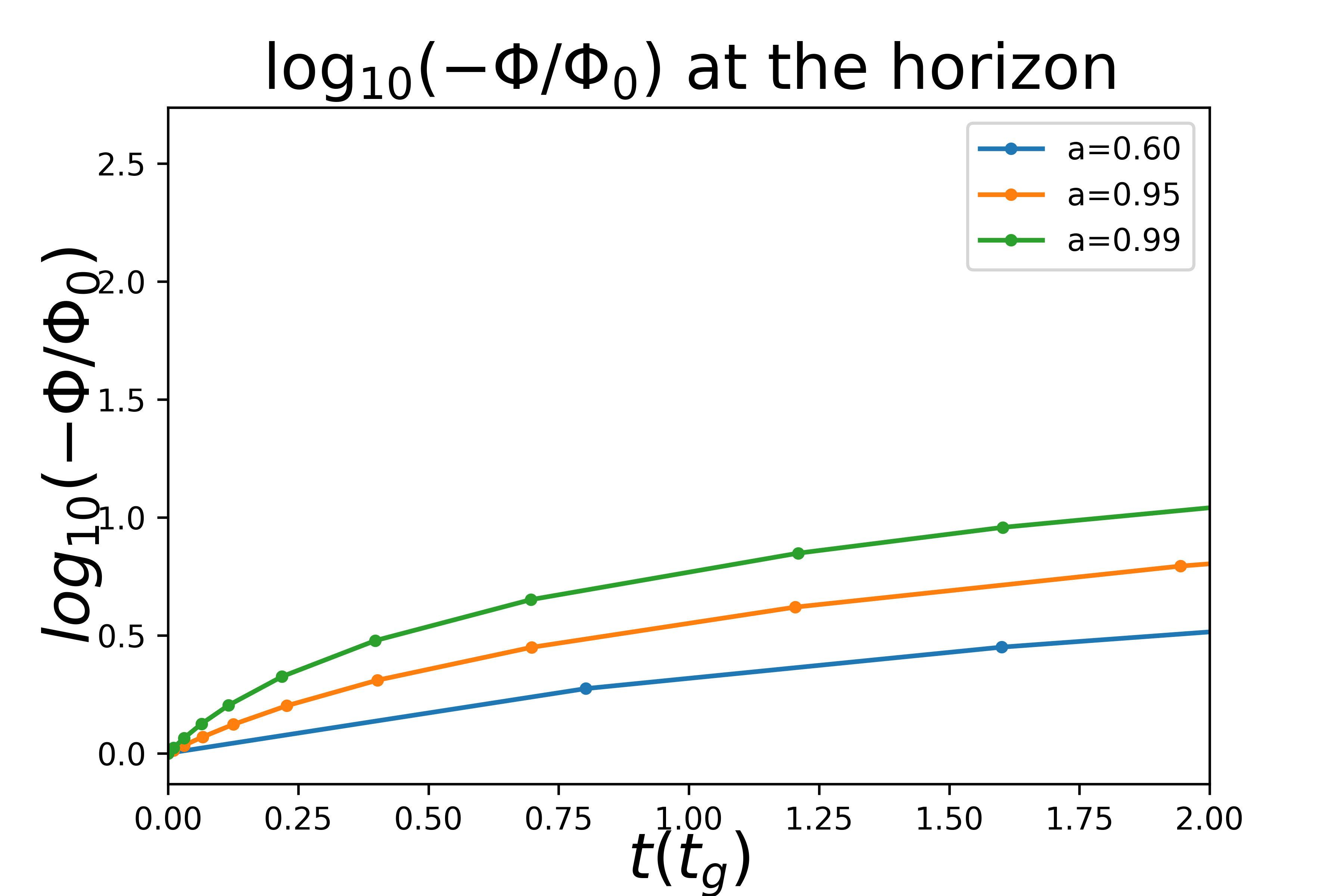}
 \caption{The initial phase of the flux evolution from the previous figure is shown in the semi-logarithmic scale and the dimensionless units (scaled by the typical values). At time zero the magnetic flux of the Wald configuration vanishes for Kerr black hole at extreme rotation ($a\rightarrow1$) in agreement with the Meissner effect. The magnetic expulsion is eliminated as soon as some plasma gets accreted, even in the purely 2D configuration with the imposed axial symmetry.}
 \label{fig5}
\end{figure}

In the strongly magnetized case the accretion rate drops because an outflow develops in the equatorial plane, where the field lines expand radially and carry plasma with them. The field line expansion is not uniform across the integration domain, leading to the distortion of the field lines and eventually to the formation of the current sheet and the reconnection events happening in the equatorial plane (notice the associated spikes around $t=180$). It is illustrative to plot the inverse accretion rate, where we can clearly identify four different phases of the system evolution (see Fig.~\ref{fig3}). The section denoted A is where the initial Bondi accretion prevails; in part B we notice the formation of the current sheet followed by a rarefaction phase of the field lines in part C. Here, more matter is expelled from the black hole and it eventually ends up in a region of much lower density. In the D phase the reconnection events occur and the outflow accelerates near the equator.

\begin{figure}[tbh!]
 \centering
 \includegraphics[width=0.49\textwidth]{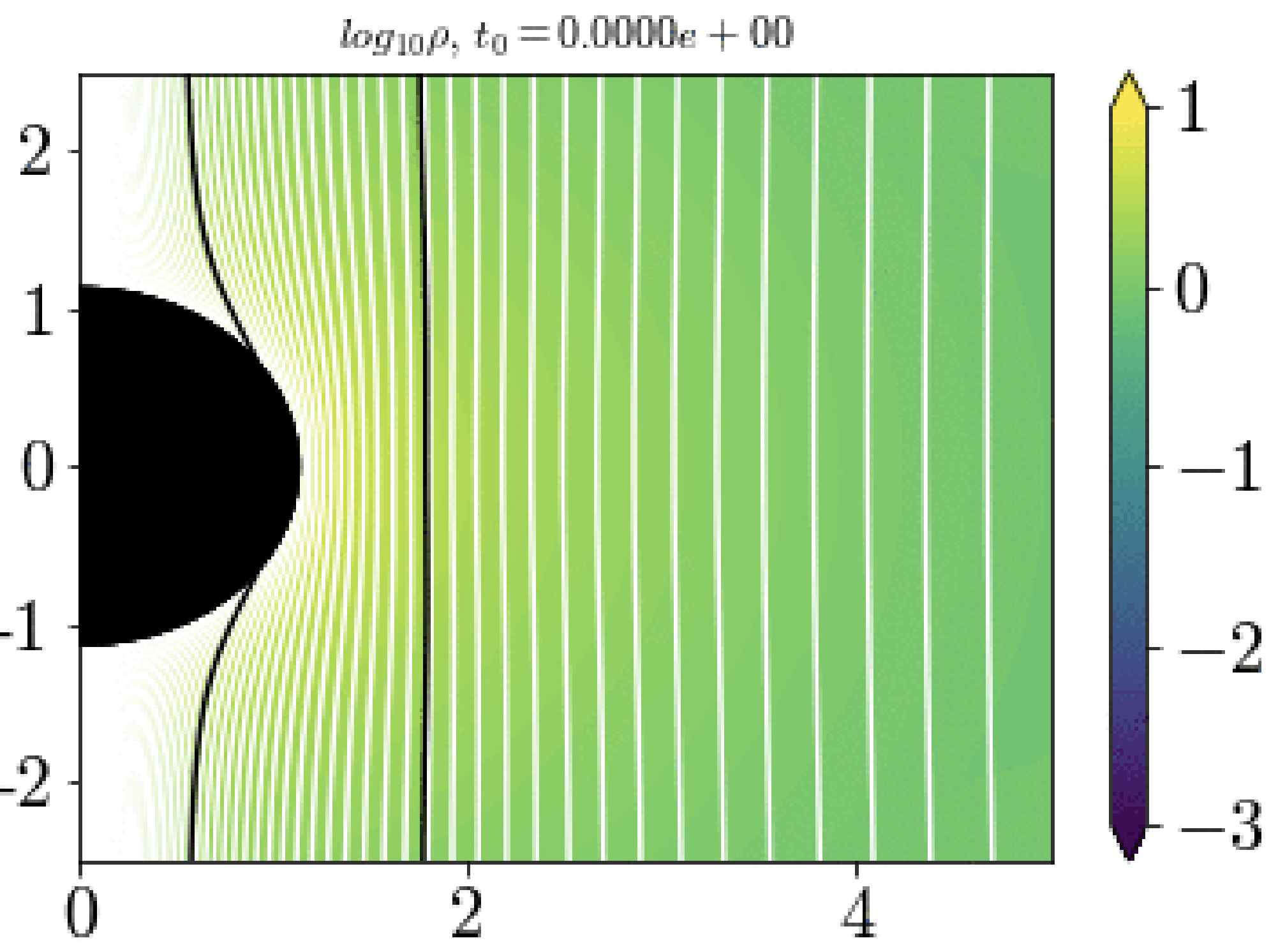}
 \includegraphics[width=0.49\textwidth]{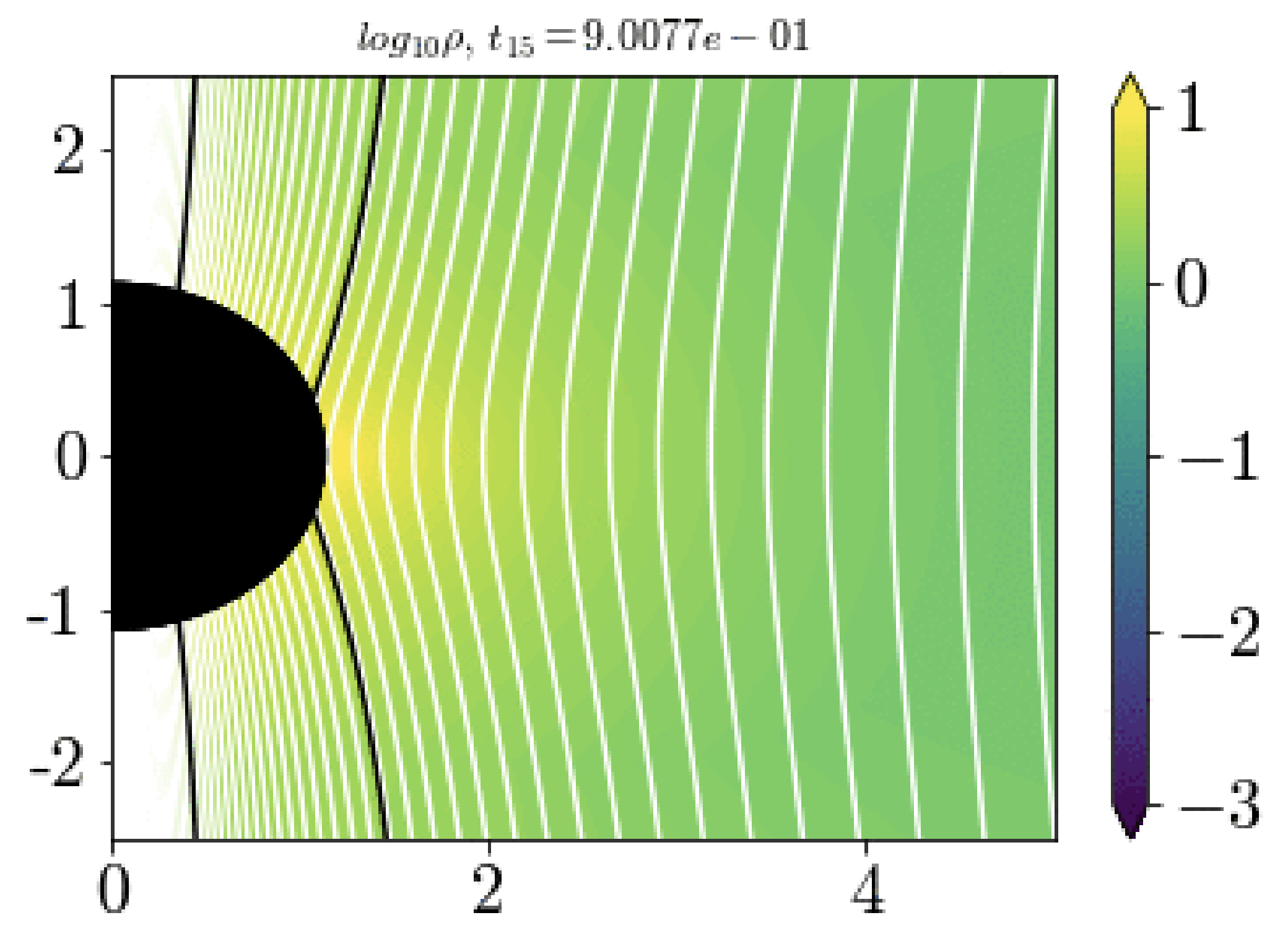}
 \includegraphics[width=0.49\textwidth]{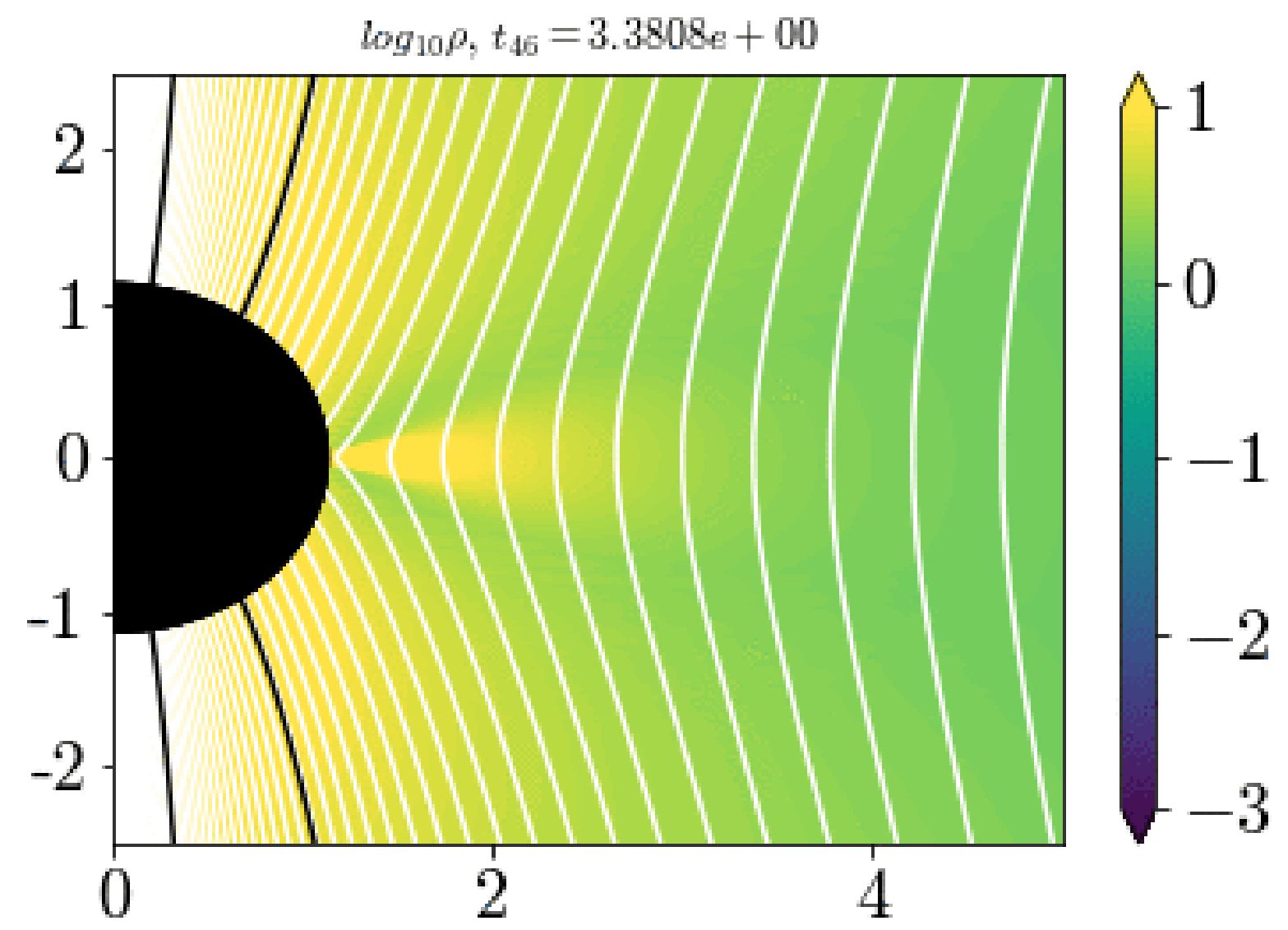}
 \includegraphics[width=0.49\textwidth]{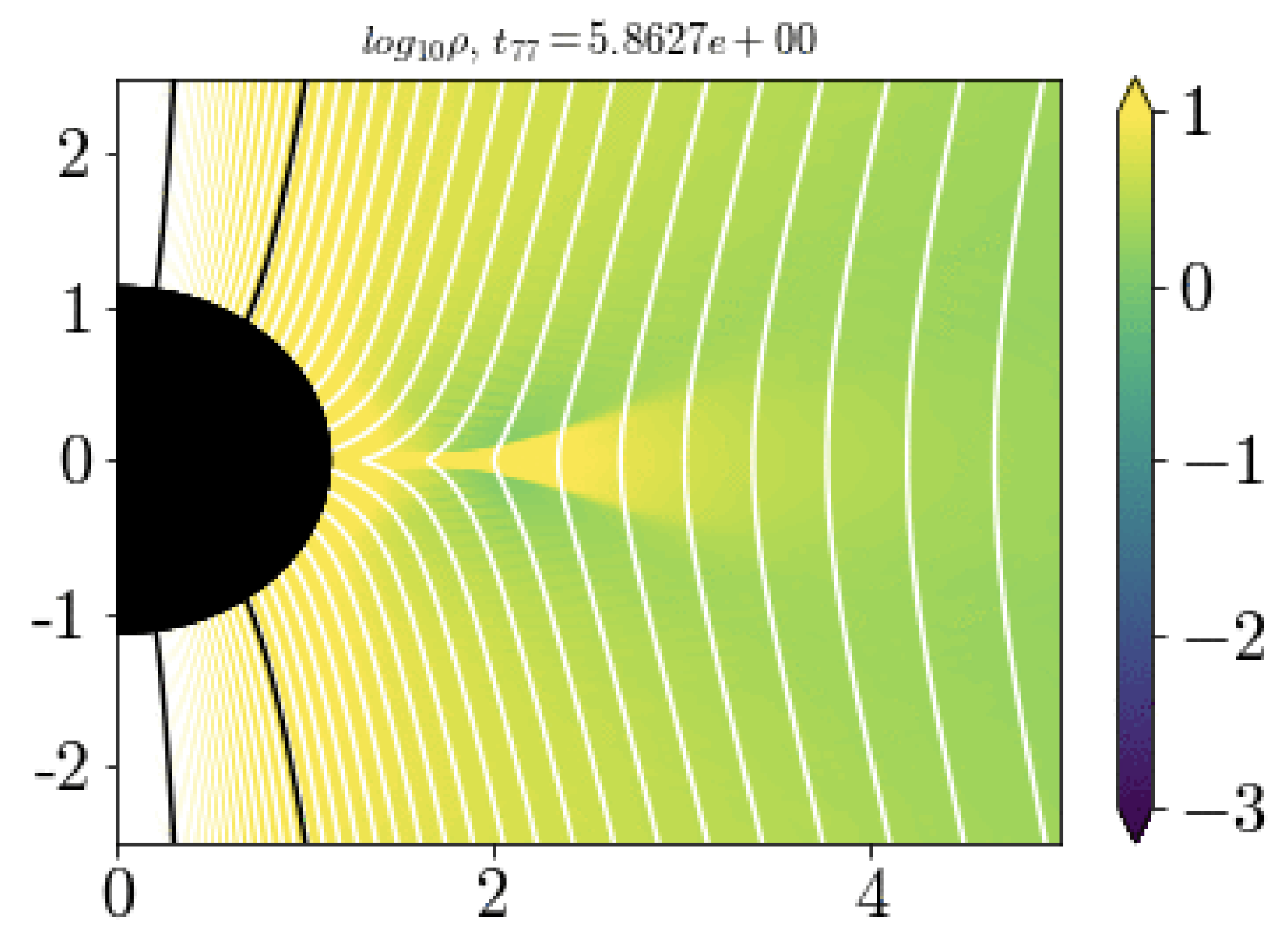}
 \caption{Four time steps (from top left to bottom right panels) with the magnetic field lines and colour-coded plasma density near a rapidly rotating ($a=0.99$) black hole with $B=8$. Magnetic field lines are expelled out of the horizon at $t=0$ (before plasma starts to be accreted), but they start crossing the horizon and accreting with the plasma as time passes. We intentionally select the magnetically dominated system at the initial stage, $\beta(t=0)\ll1$, which is expected to support the Meissner expulsion. Still, once some plasma arrives at the event horizon, we observe rapid accretion of the magnetic flux which cannot inhibit the radial motion. At late stages the field lines adopt more radial configuration near horizon and they eventually induce the ejection of plasma in the equatorial plane. The logarithmic scale of density is shown on the colour bar in arbitrary units. Time (growing in the three snapshots from left to right) corresponds to the code units.}
 \label{fig6}
\end{figure}

\begin{figure}[tbh!]
 \centering
 \includegraphics[width=0.49\textwidth]{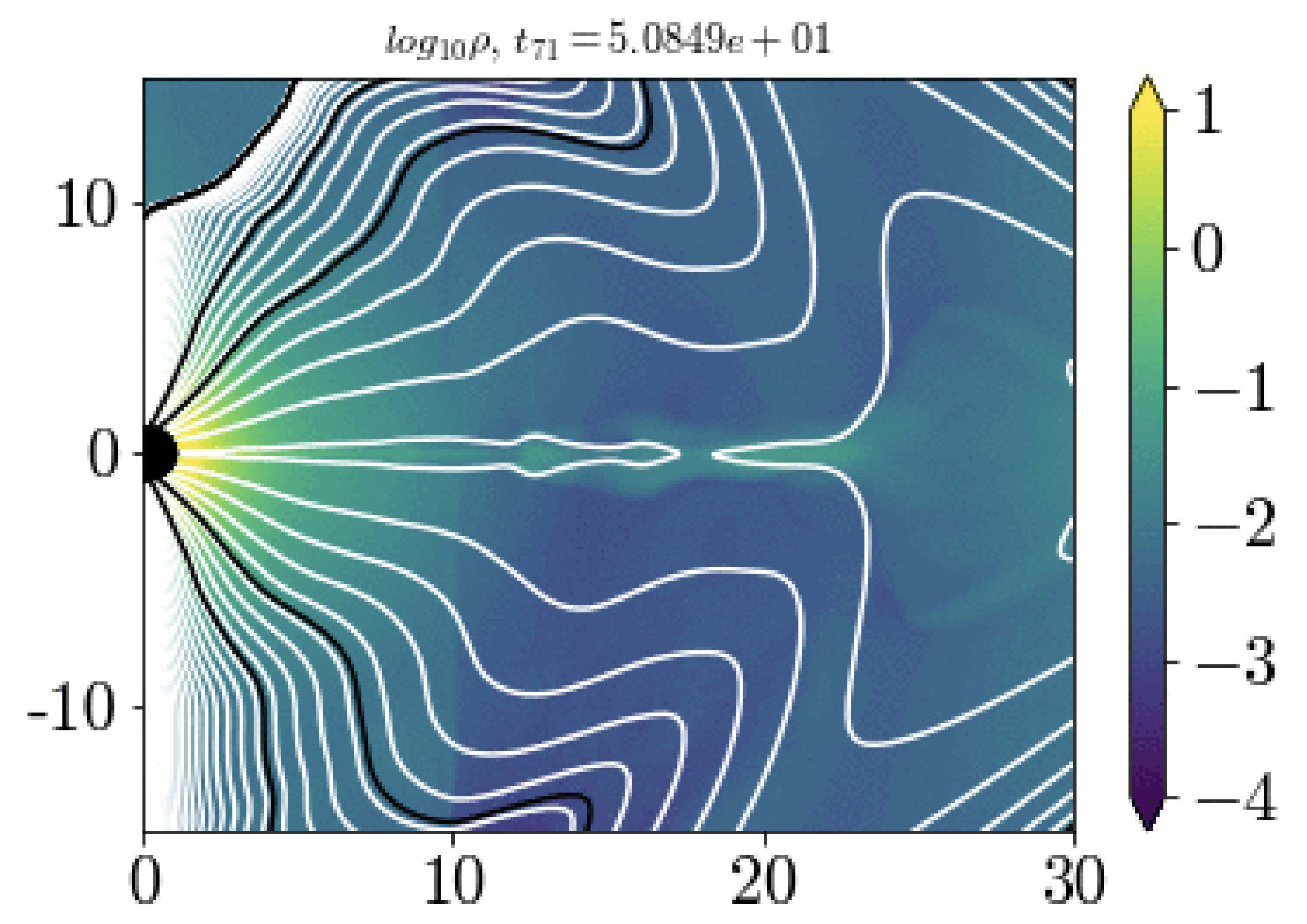}
 \includegraphics[width=0.49\textwidth]{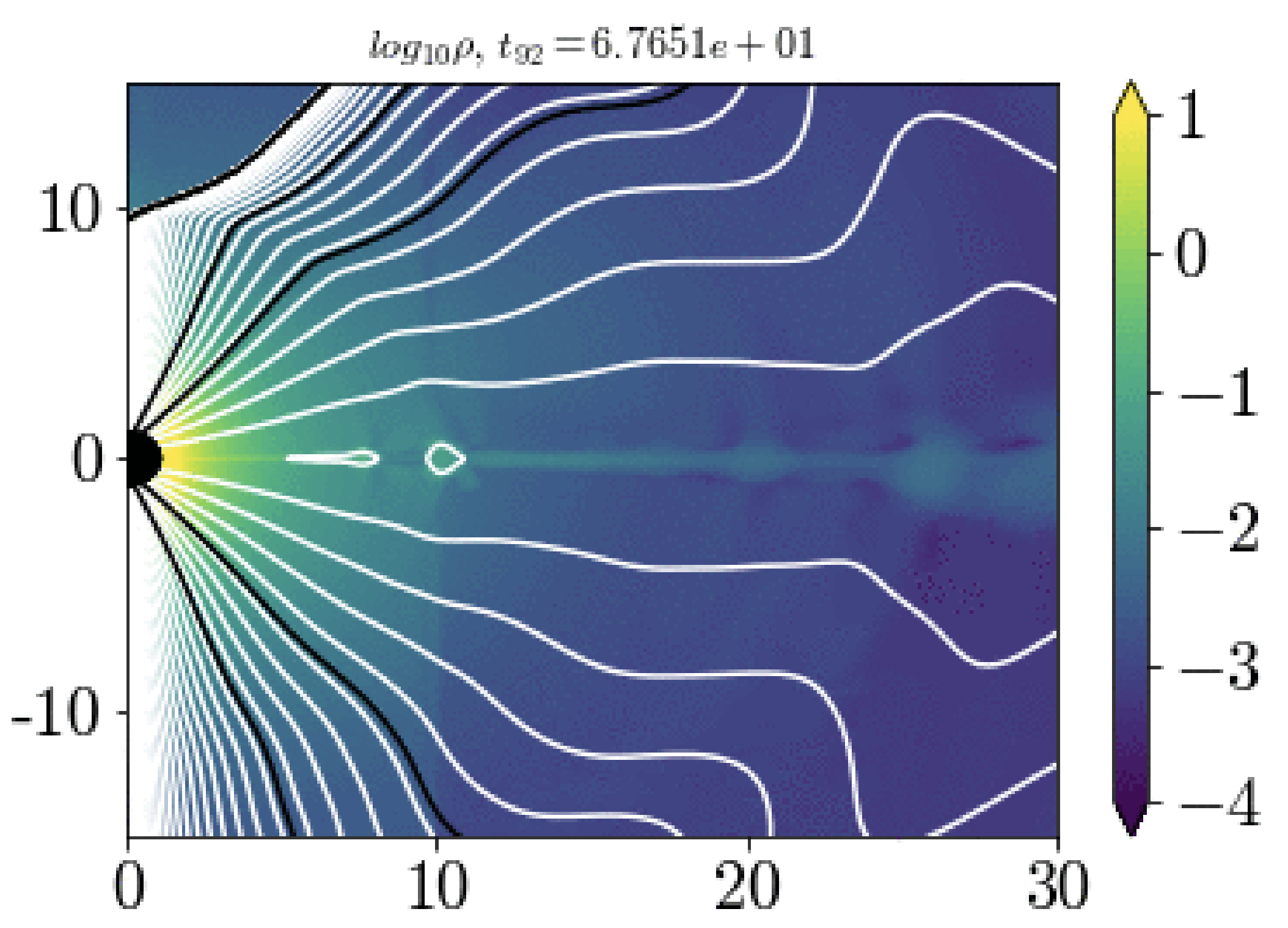}
 \includegraphics[width=0.49\textwidth]{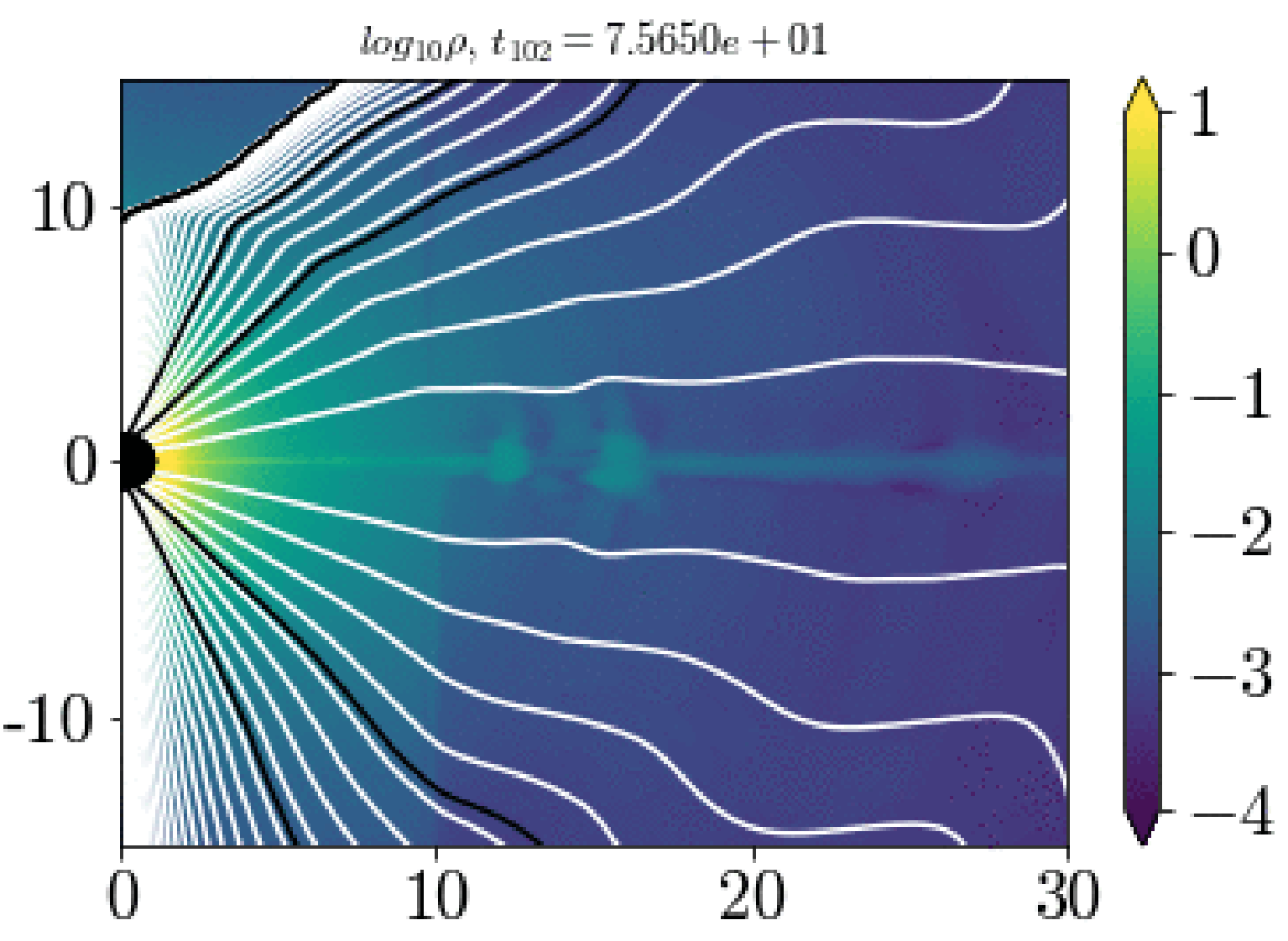}
 \includegraphics[width=0.49\textwidth]{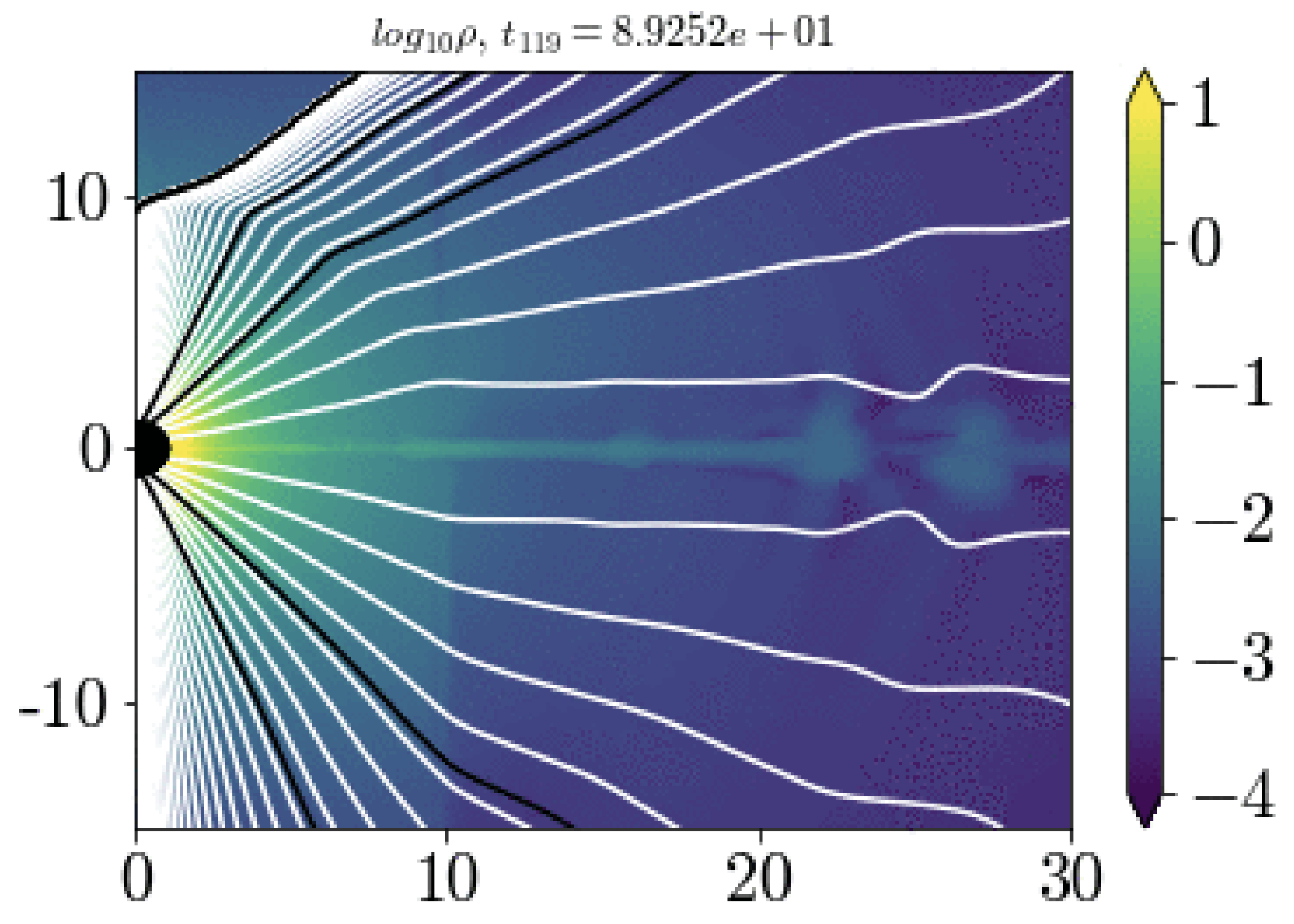}
 \caption{Similar as in the previous figure but plotted here over a larger integration domain and longer period of time. Ejection of equatorial material is observed in the form of plasmoids, which represent a disk or a ring of outflowing material in the adopted 2D approximation. In the vicinity of the rotating black hole ergosphere the frame-dragging effect acts on the plasma as well as the magnetic field lines. As a result of the frame-dragging mechanism, the outflowing material adopts a toroidal component of the orbital motion.}
 \label{fig7}
\end{figure}

\subsection{Magnetic flux across a hemisphere}
In order to reveal the changes of the magnetic field near the black hole, we study the evolution of the magnetic flux inflowing into the hemisphere located on the horizon. As mentioned above, the initial configuration is the vacuum solution of the uniform flux tube oriented in the direction parallel to the rotation axis, i.e.\ the Wald solution, but this configuration starts quickly changing once the inflowing plasma arrives in the domain.\footnote{Some more complicated configurations were explored by \citet{2020arXiv200407535K}. Interestingly, these authors found that a parabolic magnetic field also develops in the accretion torus funnel around the vertical axis, for any initial magnetic field configuration.} As also mentioned above, the initial configuration for the plasma inflow is the Bondi solution. As soon as a field line enters the ergosphere, it has to terminate at the event horizon. Therefore, the magnetic field lines in a force-free magnetosphere are not expelled by even extreme rotation of the black hole. Figure~\ref{fig4} shows the magnetic flux as a function of time and Fig.~\ref{fig5} exhibits a more detailed view of the brief initial period. Finally we vary the value of the Wald magnetic intensity $B$ parameter to obtain more initially magnetized case (lower plasma $\beta$). Notice that the code itself poses a limit $\beta>10^{-4}$ under which it starts introducing artificial density floor in order to avoid numerical integration problems. Figures \ref{fig6}--\ref{fig7} show several snapshots of the magnetic field and plasma density at different resolution. An equatorial outflow forms at late stages of the system evolution and it drives plasmoids in the outward direction.

\section{Conclusions}
Astrophysical black holes can be detected and their parameters constrained by observations in the electromagnetic domain only if the accretion process takes place, lights up the system, and produces the characteristic spectral features and variability of the emerging radiation signal. However, the properties of the cosmic environment vary in very broad range: from magnetically dominated (almost) vacuum filaments of the organized field lines to the force-free field lines frozen in the accreting medium. 

In the present contribution we were interested to explore the transition between two extreme states: from the initial configuration, where the Wald--type uniform field (aligned with the black hole rotation axis) comes to the contact with the Bondi--type radial inflow solution. The fluid drags the field lines onto the black hole, and simultaneously becomes influenced and partly expelled by the evolving magnetic field until the final state is reached after many dynamical periods. Let us note that the initial vacuum, ordered, homogeneous, parallel to the rotation axis magnetic field is an idealized situation. It has been frequently employed in order to define the starting configuration and we also use it as a test bed solution that can represent a rotating black hole embedded into a large-scale (exceeding the size of the horizon) magnetic filament, which allows us to model the rapid disappearance of the magnetic expulsion once the conducing medium starts to be accreted. An interesting development emerges as the magnetic lines are bent in the radial direction near the horizon and they start reconnecting in the equatorial plane, thus accelerating the outflow in the direction perpendicular to the rotation axis. In fact, the accretion of magnetic field lines onto the black hole starts the process of their bending from the initial Wald configuration to the split-monopole topology, which leads to the rapid disappearance of the Meissner expulsion in our system. 

{\em We can suggest that the resulting equatorial outflow is possible thanks to the fact that the Meissner effect does not operate in the magnetosphere filled with plasma.} 

Let us note that accretion disk backflows have been observed in various circumstances including the simulations of accreting black holes and stars \citep[see][]{2000astro.ph..6266K,2020arXiv200601851M}. In several aspects the system discussed in our present work is rather distinct: it does not include a magnetic star as a source of dipole-like magnetic field (we considered a black hole in the centre, which was magnetized by external currents), neither an equatorial accretion disk as the initial condition (Bondi spherical inflow was assumed as the condition at the outer boundary). Indeed, we suggest that the backflows are rather generic features that can occur in different accreting systems.

Abandoning the axial symmetry will be the next step towards a more realistic description. Also, once an oblique magnetic field (inclined with respect to the rotation axis) and a twisted (non-spherical as well as non-axisymetric) accretion flow are considered, we can expect the outflowing plasmoids to be scattered in a wide range of directions. 

\ack
The authors thank the Czech Ministry of Education Youth and Sports Mobility project No.\ 8J20PL037 to support the Czech-Polish scientific cooperation, and the Czech Science Foundation -- Deutsche Forschungsgemeinschaft collaboration project No.\ 19-01137J on processes around supermassive black holes. KS and AJ were supported by grants No.\ 2016/23/B/ST9/03114  and  2019/35/B/ST9/04000  from  the Polish National Science Center.  We acknowledge computational resources of the Warsaw ICM through the grant Gb79-9. The authors thank the referee for helpful comments and a number of suggestions to improve our text.


\end{document}